\documentclass{article}
\usepackage{graphicx}
\usepackage{todonotes}
\usepackage{listings}
\usepackage{xcolor}
\usepackage[normalem]{ulem}
\usepackage{appendix}
\usepackage{amsmath,amssymb,amsthm}
\usepackage{url}
\usepackage{float}
\usepackage{multirow}
\usepackage{algorithm}
\usepackage{algorithmic}
\usepackage[margin=1.2in]{geometry}

\usepackage{hyperref}

\usepackage{float} 

\newcommand{\edits}[1]{\textcolor{black}{#1}}

\providecommand{\keywords}[1]
{
  \small	
  \textbf{Keywords---} #1
}

\usepackage[]{biblatex}
\title{Normalizing Basis Functions: Approximate Stationary Models for Large Spatial Data}

\author{Antony Sikorski\footnote{Corresponding author: asikorski@mines.edu} , Daniel McKenzie, and Douglas Nychka}
\date{Department of Applied Mathematics and Statistics, Colorado School of Mines}

\begin{document}

\maketitle

\begin{abstract}
    In geostatistics, traditional spatial models often rely on the Gaussian Process (GP) to fit stationary covariances to data. It is well known that this approach becomes computationally infeasible when dealing with large data volumes, necessitating the use of approximate methods. A powerful class of methods approximate the GP as a sum of basis functions with random coefficients. Although this technique offers computational efficiency, it does not inherently guarantee a stationary covariance. To mitigate this issue, the basis functions can be “normalized” to maintain a constant marginal variance, avoiding unwanted artifacts and edge effects. This allows for the fitting of nearly stationary models to large, potentially non-stationary datasets, providing a rigorous base to extend to more complex problems. Unfortunately, the process of normalizing these basis functions is computationally demanding. To address this, we introduce two fast and accurate algorithms to the normalization step, allowing for efficient prediction on fine grids. The practical value of these algorithms is showcased in the context of a spatial analysis on a large dataset, where significant computational speedups are achieved. While implementation and testing are done specifically within the LatticeKrig framework, these algorithms can be adapted to other basis function methods operating on regular grids. 
    
\end{abstract}
\vspace{0.1in}
\keywords{Spatial Statistics, Basis Function Models, Kriging, Gaussian Process, Big Data, Stationary Process, FFT, LatticeKrig}
\vspace{0.1in}

\section{Introduction}
\label{sec:Intro}
Generating prediction surfaces in geostatistics frequently involves fitting models with a stationary covariance to spatially referenced data on a regular grid and then computing model predictions at many unobserved grid points. Historically, the Gaussian Process (GP) has served as a primary model for such analysis, as GPs allow for straightforward prediction and uncertainty quantification \cite{cressie2015statistics, schabenberger2017statistical}. Well known computational bottlenecks arise from the need to evaluate the conditional expectation to obtain a spatial prediction, and the likelihood to estimate covariance parameters. These operations generically require \(\mathcal{O}(N^3)\) operations and \(\mathcal{O}(N^2)\) memory, where \(N\) represents the number of spatial locations \cite{sungeostat, stein2008modeling}.

A flexible and computationally efficient approach for avoiding these bottlenecks is provided by a class of ``basis function models'', which approximate a spatial process as a weighted sum of basis functions \cite{cressie2022basis}. Some notable examples include Fixed Rank Kriging (FRK) \cite{frk, frkv2}, INLA \cite{INLA}, mgcv \cite{mgcv}, Multi-Resolutional Approximation (MRA) \cite{katzfussMRA}, Predictive Processes (PP) \cite{banerjee2008gaussian, finley2009improving}, Spatial Partitioning \cite{heatonspatialpartition}, and LatticeKrig \cite{latticekrig, LKrigpackage}. Despite variations in basis function choice and modeling approaches, a common disadvantage is the absence of guaranteed stationarity in the covariance structure, which mandates that the spatial dependence between two points rely solely on their distance. Although data are frequently non-stationary, well fitting stationary models allow for easier and more efficient parameter estimation due to the simplicity of the covariance structure, along with the use of standard estimation methods. Moreover, non-stationary models are often built by modifying a stationary covariance kernel. 

\begin{figure}[h]
    \centering
    \includegraphics[width = 1\textwidth]{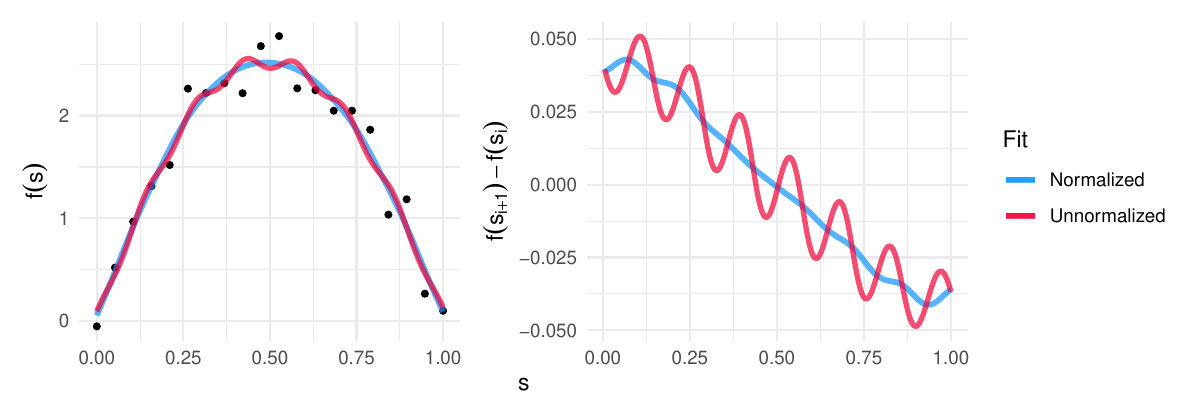}
    \caption{ \small A simple, one-dimensional scenario, where a small amount of noise is added to data sampled from a quadratic. {\bf Left:} Both normalized and un-normalized basis function models are fitted to the data. {\bf Right:} Finite difference approximation of the gradient for both fits reveals the artifacts more clearly.}
    \label{fig:1d-patterns}
\end{figure}

A drawback of basis function methods is the presence of undesirable artifacts in the predicted surface, as demonstrated in Figure \ref{fig:1d-patterns}. Here a simple model consisting of eight overlapping basis functions is fit to noisy observations from a quadratic function. A naive fit introduces oscillations in the predicted curve that are solely due to the placement and shape of the basis functions. The reduction or removal of these artifacts is critical to achieving stationarity, and fitting models that are representative of the physical data. One way of doing so is by ``normalizing'' the basis functions to ensure a constant marginal variance.   

\noindent Although it is an improvement to the spatial model, this normalization step constitutes a subsequent computational bottleneck. In order to analyze datasets on the order of millions or more spatial locations ($N \sim 10^6$) efficiently, it has become necessary to explore, understand, and accelerate this operation.  

In this paper, we significantly reduce the computational cost of ensuring the stationarity of basis function models on a regular grid. We provide two algorithms, and software for reproducibility. The first algorithm uses the fast Fourier transform (FFT) to approximate the variance of the basis functions on large grids rapidly. This replaces an $\mathcal{O}(N^3)$ exact computation with an $\mathcal{O}(N\log(N))$ approximate one, at almost no loss of accuracy. The second algorithm calculates the variance exactly by decomposing key matrices into a series of Kronecker products of smaller matrices. All supplementary code for the figures and experiments in this paper, along with a development version of the {\tt LatticeKrig} R package can be found on \href{https://github.com/antonyxsik/Normalization-Paper}{\tt github.com/antonyxsik/Normalization-Paper}. 

\subsection{Paper Outline}
\label{sec:Outline}

The remainder of this paper is as follows. Section~\ref{sec:BFMethods} provides a brief introduction to basis function methods, and as a concrete example, the LatticeKrig model. In Section~\ref{sec:Methods}, two fast normalization methods are described: one based on discretization using Kronecker matrices and the other based on image upsampling using the Fast Fourier Transform (FFT). Section~\ref{sec:Results} provides experimental timing and accuracy results. Section~\ref{sec:Application} demonstrates the practical value of the algorithms during the analysis of a large, simulated dataset, while conclusions and recommendations for future work are contained in Section~\ref{sec:Conclusion}.

\section{Basis Function Methods for Spatial Prediction}
\label{sec:BFMethods}

A ubiquitous model for a spatial process $Z$ is 
\begin{equation}
    Z(\mathbf{s}) = Y(\mathbf{s}) + \varepsilon(\mathbf{s}) 
\end{equation}
where $Z(\mathbf{s})$ is the observation at spatial location $\mathbf{s}$, $Y(\cdot)$ is a Gaussian Process, and $\varepsilon(\cdot)$ encapsulates \edits{independent, normally distributed measurement error with mean zero and a variance of $\tau^2$.} We typically assume that the mean of $Y(\cdot)$ depends linearly on a number of observed covariates $x_1(\cdot), \ldots x_d(\cdot)$ 
\begin{equation}
    Y(\mathbf{s}) = \mathbf{x}(\mathbf{s})^{\top}\boldsymbol{\beta} + g(\mathbf{s}), \quad \text{ where } \mathbf{x}(\mathbf{s}) = \begin{bmatrix} x_1(\mathbf{s}) & \cdots & x_d(\mathbf{s})\end{bmatrix},
\end{equation}
although more sophisticated models are possible \edits{\cite{zhan2023neural, saha2023random, sigrist2022gaussian}}. Note that $g(\cdot)$ is a zero-mean Gaussian Process; we denote its covariance function as $C(\cdot, \cdot)$:
\begin{equation}
    \mathrm{cov}\left(g(\mathbf{s}), g(\mathbf{s}^{\prime})\right) = C(\mathbf{s},\mathbf{s}^{\prime}).
\end{equation}

A central problem in this framework is to predict $Y(\mathbf{s})$ at large numbers of unobserved locations. We focus on the case where these locations lie on a square grid\footnote{Our methods apply equally to rectangular grids; we restrict to square grids in our exposition for notational clarity} $\{\mathbf{s}_{i,j}\}_{i,j=1}^{n}$. Fitting a model for this purpose with standard frequentist methodology involves parameter estimation ({\em i.e.} computing $\boldsymbol{\beta} \in \mathbb{R}^{d}$ and any free parameters in $C(\cdot, \cdot)$) by maximum likelihood. These computations are dominated by evaluating the inverse and determinant of matrices of size $N \times N$, where $N = n \times n$, with a generic cost of $\mathcal{O}(N^3)$ operations and $\mathcal{O}(N^2)$ memory. For small problems this framework has the advantage that any covariance kernel can be considered to model spatial dependence in $Y$. However, these computations become prohibitive for even moderately sized datasets ({\em i.e.} $N\sim 10^4$).

Basis function models, also known as fixed-rank approaches, sidestep this computational barrier by either reducing the size of key matrices or introducing sparsity into the model. We focus on approaches in which a set of $R = r \times r$ basis functions $\{\varphi_{k,l}(\cdot)\}_{k,l = 1}^r$ is constructed via translations and dilations of a fixed, compactly supported, unimodal function $\psi$:
\begin{equation}
    \varphi_{k,l}(\mathbf{s}) = \psi\left( \left\| \frac{\mathbf{s} -\mathbf{u}_{k,l}}{\gamma} \right\| \right).
\end{equation}
Here $\gamma$ is a scaling factor that controls the overlap of the basis functions. Common choices for $\psi$ include bisquare functions \cite{frk}, Wendland polynomials \cite{latticekrig}, and wavelets \cite{vidakovic1999wavelets}. We further assume that the {\em basis function centers} $\{\mathbf{u}_{k,l}\}_{k,l=1}^r$ form a square, regular grid. Basis function models assume an expansion for $g$ of the form 
\begin{equation}
    g(\mathbf{s}) = \sum_{k,l=1}^{r} c_{k,l}\varphi_{k,l}(\mathbf{s}).
    \label{eq:basis_function_model}
\end{equation}
Randomness and correlation in $g(\cdot)$ is now captured by the coefficients $c_{1,1},\ldots, c_{r,r}$. We assume $\mathbf{c}$ is drawn from a multivariate Gaussian, $\mathbf{c} \sim \mathcal{MN}(0, \Sigma)$, 
 with covariance matrix $\Sigma \in \mathbb{R}^{R \times R}$. We define $\Phi \in \mathbb{R}^{N\times R}$ as the {\em regression matrix}, with columns indexing the basis functions and rows indexing locations. The full ``Kriging'' prediction for this model is then given by
\begin{equation}
    \hat{Y}(\mathbf{s}) = \mathbf{x}(\mathbf{s})^{\top}\hat{\boldsymbol{\beta}} + \hat{g}(\mathbf{s}), \quad \text{where  } \hat{g}(\mathbf{s}) = \Phi \hat{\mathbf{c}}
\end{equation} 
The coefficients $\hat{\boldsymbol{\beta}}$ are found using generalized least squares (GLS), and 
\begin{equation}
    \hat{\mathbf{c}}=(\Phi^{\top} \Phi + \edits{\tau^2} \Sigma^{-1})^{-1} \Phi^{\top}(Z-X \hat{\beta})
\end{equation}
denotes our estimate for the basis function coefficients. These forms are derived from the usual form of the spatial predictions using the Sherman-Morrison-Woodbury identity. When $\Sigma$ is unstructured, the computational costs in this model reduce from $\mathcal{O}(N^3)$ to $\mathcal{O}(NR^2)$ \cite{frk}, with $R$ chosen to be small relative to $N$. A greater number of basis functions intuitively provides a better approximation, yielding a tradeoff between computational complexity and accuracy. Reduced cost has allowed these models to be applied to numerous environmental problems with $N \sim 10^5$ or greater \edits{ \cite{frk,katzfussMRA, ma2020fused}}. 

If $\psi$ has compact support, so will $\{ \varphi_{k,l}\}$, thus $\gamma$ can be chosen such that $\Phi$ and its {\em Gram matrix} $\Phi^T \Phi$ are sparse. This motivates a final elaboration of the basis function model. Rather than computing $\Sigma^{-1}$, one can directly prescribe $Q =\Sigma^{-1}$ as a sparse precision matrix. This strategy is used in LatticeKrig \cite{latticekrig} and other popular spatial models stemming from the stochastic partial differential equation (SPDE) formulation \cite{SPDE}. Although it will later be shown that in certain settings, one must evaluate the inverse of $Q$, its structured form allows computation to be sped up using sparse linear algebra. 

Finally, we note that in practice a {\em multi-layered} approach is favored, in which basis functions centered on multiple nested, increasingly fine grids are used. For simplicity we limit our initial exposition to the single grid case, but note that our techniques directly extend to such {\em multi-resolution} approaches. 

\subsection{Prediction Artifacts in Basis Function Models}
\label{sec:PredArtifacts}
\begin{figure}[h]
    \centering
    \includegraphics[width = 0.8\textwidth]{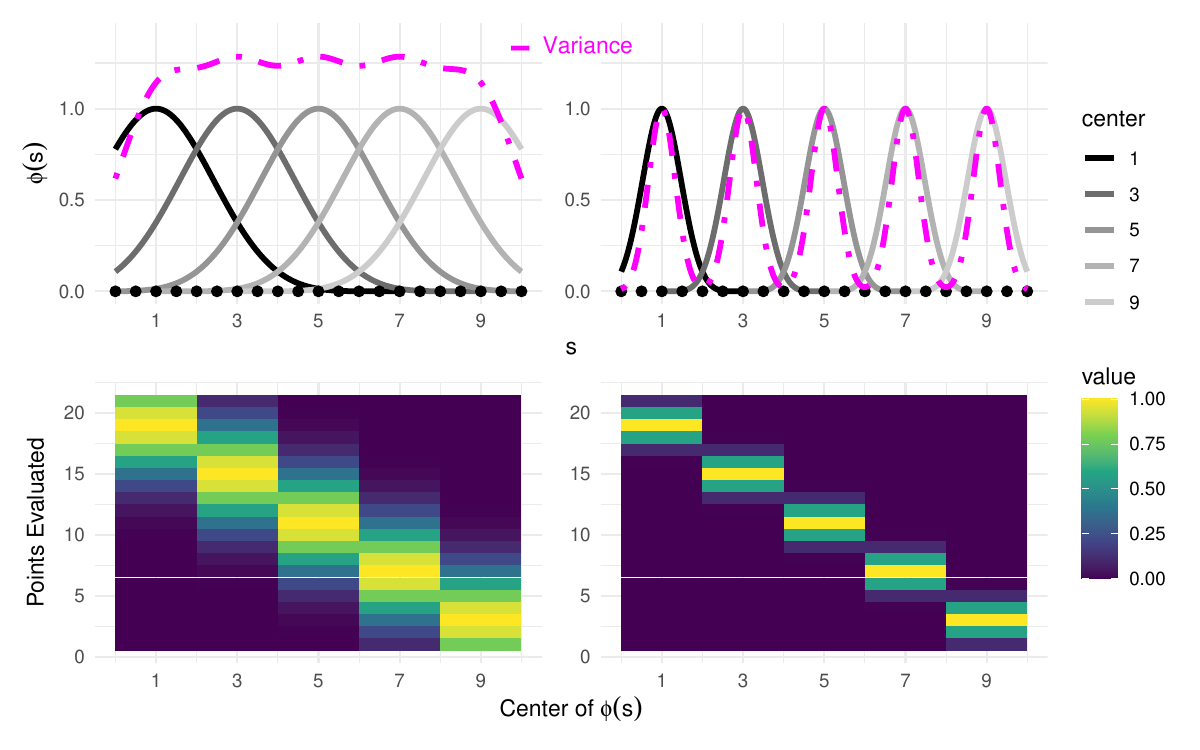}
    \caption{\small {\bf Top row:} $\varphi_1(s),\ldots, \varphi_5(s)$ where $\psi$ is a Wendland polynomial, dilated to yield large overlap ({\bf left}) and small overlap ({\bf right}). {\bf Bottom row:} the corresponding $\Phi \in \mathbb{R}^{21\times 5}$, demonstrating the sparsity pattern. While less overlap results in a sparser $\Phi$ ({\bf Right column}), it also results in a less uniform variance (magenta dotted line in {\bf Top row}).}
    \label{fig:sparsity-variance}
\end{figure}

Although basis function models can be computationally efficient, they can also produce undesirable artifacts in their prediction surfaces. Figure \ref{fig:sparsity-variance} illustrates how the choice of basis function overlap creates a trade-off between the sparsity of $\Phi$ and the uniformity of the process marginal variance. Although a sparser $\Phi$ is desirable as it reduces computational cost, the features in the marginal variance induce similar patterns in the resulting predictions.  

A natural solution to this issue is to \textit{normalize} the basis functions to guarantee a constant process variance. We define $\boldsymbol{\phi}_{\mathbf{s}} = \begin{bmatrix} \varphi_{1,1}(\mathbf{s}) & \cdots & \varphi_{r,r}(\mathbf{s})\end{bmatrix}^{\top}$ as the vector formed by evaluating all basis functions at $\mathbf{s}$. Accordingly, the basis functions are redefined as 
\begin{equation}
    \varphi_{k,l}^*(\mathbf{s})=\frac{\varphi_{k,l}(\mathbf{s})}{\sqrt{\text{Var}(g(\mathbf{s}))}}, \quad \text{where } \text{Var}(g(\mathbf{s})) = \boldsymbol{\phi}_{\mathbf{s}}^{\top} \Sigma \boldsymbol{\phi}_{\mathbf{s}}
\end{equation}
It has previously been shown that this transformation provides a good approximation of a stationary GP \cite{latticekrig}. For methods that define $\Sigma$ explicitly, this computation is straightforward. However, for methods that are phrased in terms of the precision matrix this normalization is problematic. Here the variance computation is 
\begin{equation}
    \text{Var}(g(\mathbf{s})) = \boldsymbol{\phi}_{\mathbf{s}}^{\top} Q^{-1} \boldsymbol{\phi}_{\mathbf{s}}
\end{equation}
which requires evaluating $Q^{-1}$. Note that the normalization as proposed requires evaluation not only at all observed data locations, but also at all locations for prediction. When generating large prediction surfaces or predicting on fine grids, this becomes a bottleneck for otherwise very fast methods.

\subsection{LatticeKrig Spatial Model} 
\label{sec:LK}
The LatticeKrig (LK) methodology shares many elements with earlier frameworks such as FRK, but directly prescribes a sparse representation of $Q$ to allow for fast calculations. The resulting model allows for the total number of basis functions $R$ to exceed the number of spatial locations $N$, with the limiting factor being the sparsity of the matrix $(\Phi^{\top} \Phi+ \edits{\tau^2} \Sigma^{-1})$ rather than its $R \times R$ nominal size. This allows for fitting spatial data where the measurement error is small and $g$ is close to interpolating the observations. 

Basis function coefficients $\mathbf{c}$ are defined implicitly as a spatial autoregression (SAR), such that $B\mathbf{c} = \mathbf{e}$, where $B$ is a sparse matrix and $\mathbf{e}$ are independent $\mathcal{N}(0,1)$ random variables. For two dimensions, the diagonal elements of $B$ are $4 + \kappa^2$, while the four nearest neighboring grid points are $-1$. The $\kappa^2$ parameter is adjustable, and defines the correlation range of the SAR model. For interior grid centers this results in the corresponding row of $B$ having only 5 non-zero values. Linear statistics then implies that $Q = BB^{\top}$ is the precision matrix of the coefficients $\mathbf{c}$.

Although the LK model enjoys many benefits, the direct prescription of the precision matrix makes normalization problematic, as evaluating $Q^{-1}$ is required for the variance calculation. Since $Q$ is positive definite, the default method to calculate $\mathrm{Var}(g(\mathbf{s}))$ in LK first computes the sparse Cholesky decomposition $Q = DD^T$ and then solves $D\mathbf{v} = \boldsymbol{\phi}_{\mathbf{s}}$, whence 
\begin{equation}
\mathrm{Var}(g(\mathbf{s})) = \|\mathbf{v}\|_2^2 =  \sum_{i,j=1}^{r} v_i^2. 
\label{eq:var_as_norm}
\end{equation}

In this work we propose a different approach, based on an alternative factorization of $Q$ that is readily available, namely $Q = BB^{\top}$ where $B$ is the SAR matrix described above. In fact, $B$ can be decomposed into $B = L + \kappa^2I$ where $L$ is the five-point stencil discrete Laplacian \cite{burden1997numerical}. Thus we propose to solve
\begin{equation}
   (L + \kappa^2 I)\mathbf{v} = \boldsymbol{\phi}_{\mathbf{s}}
   \label{eq:exact_variance_solve}
\end{equation}
instead, and then compute $\mathrm{Var}(g(\mathbf{s}))$ as the norm of $\mathbf{v}$, as in \eqref{eq:var_as_norm}. Exploiting the structure of \eqref{eq:exact_variance_solve} provides multiple avenues to speed up the computation of $\mathrm{Var}(g(\mathbf{s}))$.

\section{Methods}
\label{sec:Methods}

\subsection{Normalization using the FFT}

In deriving this method, we were inspired by the observation that the variance function $\mathbf{s} \mapsto \mathrm{Var}(g(\mathbf{s}))$ appears nearly periodic. Before continuing, we offer a justification of why this is the case.

\subsubsection{Approximate periodicity of the variance function}
\label{section:Justifying_Periodicity}
Because the $\varphi_{k,l}$ are constructed as grid translations of a single parent function $\psi$, 
\begin{align*}
    \boldsymbol{\phi}_{\mathbf{s}} &= \begin{bmatrix} \varphi_{1,1}(\mathbf{s}) & \ldots & \varphi_{r,r}(\mathbf{s}) \end{bmatrix} \\
    & = \begin{bmatrix} \psi(\|\mathbf{s} - \mathbf{s}_{1,1}\|/\gamma) & \ldots & \psi(\|\mathbf{s} - \mathbf{s}_{r,r}\|/\gamma) \end{bmatrix} \\
    & =\begin{bmatrix} \psi_{\mathbf{s}}(\mathbf{s}_{1,1}) & \ldots & \psi_{\mathbf{s}}(\mathbf{s}_{r,r})\end{bmatrix}.
\end{align*}

To simplify matters and avoid edge effects, we consider the case where the basis function centers form an infinite grid, specifically the integer lattice $\mathbb{Z}^2$. In this case, the variance is
\begin{equation}
    \mathrm{Var}(g(\mathbf{s})) = \sum_{\mathbf{m}\in\mathbb{Z}^2}v_{\mathbf{m}}^2,
    \label{eq:lattice_variance}
\end{equation}
where we ignore---for now---issues of summability. The linear system of interest is now a difference equation that is a discretization of a well known partial differential equation (PDE), the {\em screened Poisson equation},
\begin{equation}
    -\Delta v_{\mathbf{s}} +\kappa^2 v_{\mathbf{s}} = \psi_{\mathbf{s}}(\mathbf{z}). 
    \label{eq:screened_poisson}
\end{equation}
where the solution $v_{\mathbf{s}}:\mathbb{R}^2\to\mathbb{R}$ depends on the source term $\psi_{\mathbf{s}}(\mathbf{z})$. The PDE \eqref{eq:screened_poisson} may be solved using the Fourier transform, as discussed further in \edits{Appendix \ref{appendix_a}}, yielding the solution
\begin{equation}
    v_{\mathbf{s}}(\mathbf{z}) = \iint_{\mathbb{R}^2}\frac{e^{-2\pi i(\mathbf{s} - \mathbf{z})\cdot\boldsymbol{\omega}}}{\kappa^2 + \|\boldsymbol{\omega}\|^2}\tilde{\psi}(\boldsymbol{\omega})d\boldsymbol{\omega},
\end{equation}
where $\tilde{\psi}(\cdot)$ denotes the Fourier transform of $\psi$.  
Returning to \eqref{eq:lattice_variance}:
\begin{equation}
    \mathrm{Var}(g(\mathbf{s})) = \sum_{\mathbf{m}\in\mathbb{Z}^2}|\mathbf{v}_{\mathbf{s}}(\mathbf{m})|^2 = \sum_{\mathbf{m}\in\mathbb{Z}^2} \iint_{\mathbb{R}^2}\frac{e^{-2\pi i(\mathbf{s} - \mathbf{m})\cdot\boldsymbol{\omega}}}{\kappa^2 + \|\boldsymbol{\omega}\|^2}\tilde{\psi}(\boldsymbol{\omega})d\boldsymbol{\omega}
\end{equation}
it is now apparent that $\mathrm{Var}(g(\mathbf{s}))$ is $\mathbb{Z}^2$-periodic, as for any $\mathbf{n}\in\mathbb{Z}^2$
\begin{align}
    \mathrm{Var}(g(\mathbf{s} + \mathbf{n})) &= \sum_{\mathbf{m}\in\mathbb{Z}^2}|\mathbf{v}_{\mathbf{s} + \mathbf{n}}(\mathbf{m})|^2 = \sum_{\mathbf{m}\in\mathbb{Z}^2} \iint_{\mathbb{R}^2}\frac{e^{-2\pi i(\mathbf{s} + \mathbf{n} - \mathbf{m})\cdot\boldsymbol{\omega}}}{\kappa^2 + \|\boldsymbol{\omega}\|^2}\tilde{\psi}(\boldsymbol{\omega})d\boldsymbol{\omega} \\
    &= \sum_{\mathbf{m}^{\prime}\in\mathbb{Z}^2} \iint_{\mathbb{R}^2}\frac{e^{-2\pi i(\mathbf{s} - \mathbf{m}^{\prime})\cdot\boldsymbol{\omega}}}{\kappa^2 + \|\boldsymbol{\omega}\|^2}\tilde{\psi}(\boldsymbol{\omega})d\boldsymbol{\omega} \qquad \text{ where } \mathbf{m}^{\prime} := \mathbf{m} - \mathbf{n} \\
    &= \mathrm{Var}(g(\mathbf{s})),
\end{align}
and indeed we approximately observe this periodicity in practice. We conjecture that the lack of exact periodicity in real applications is primarily due to the edge effects arising from using a finite lattice, which can be reduced by adding an additional buffer region with minimal computational overhead. 
  
\subsubsection{FFT-based upsampling} 
\label{sec:FFT_method}
Based on the argument in Section~\ref{section:Justifying_Periodicity}, we expect $\mathrm{Var}(g(\mathbf{s}))$ to be nearly periodic, thus we propose to apply a commonly used image upsampling technique---{\em Fourier interpolation} \cite{sathyanarayana1990interpolation}---to approximately evaluate $\mathrm{Var}(g(\mathbf{s}))$ on the square\footnote{This also applies to rectangular grids, but we restrict to square grids in our exposition for notational clarity} grid $\{\mathbf{s}_{i,j}\}_{i,j=1}^{n}$. Fourier interpolation begins by {\em subsampling}, that is evaluating $\mathrm{Var}(g(\mathbf{s}))$ on a coarser grid which contains the grid of basis function centers: 
\begin{equation}
    \{\mathbf{u}_{k,l}\}_{k,l=1}^{r} \subset \{\mathbf{s}_{i_{p},j_q}\}_{p,q=1}^{\tilde{n}}\subset \{\mathbf{s}_{i,j}\}_{i,j=1}^{n}
\end{equation}
 and $\tilde{n} < n$. In the ideal case where $\mathrm{Var}(g(\mathbf{s}))$ is periodic, as discussed in Section~\ref{section:Justifying_Periodicity}, we observe that it is naturally {\em band-limited}, with the spacing between the $\mathbf{u}_{k,l}$ determining the highest frequency. Consequently, by the \edits{Shannon-Nyquist sampling theorem \cite[Appendix C.1]{foucart2013mathematical}}, as long as \edits{$\tilde{n} \geq 2r+1$}, the set of evaluations $\{\mathrm{Var}(g(\mathbf{s}_{i_{p},j_q}))\}_{p,q=1}^{\tilde{n}}$ determine $\mathrm{Var}(g(\mathbf{s}))$ exactly. More precisely, letting $\{\tilde{f}_{p,q}\}_{p,q=1}^{\tilde{n}}$ denote the discrete Fourier transform (DFT) of $\{\mathrm{Var}(g(\mathbf{s}_{i_{p},j_q}))\}_{p,q=1}^{\tilde{n}}$:
 \begin{equation}
     \tilde{f}_{p,q} = \sum_{p=1}^{\tilde{n}}\sum_{q=1}^{\tilde{n}}\mathrm{Var}(g(\mathbf{s}_{i_{p},j_q})) e^{-2\pi i p/\tilde{n}} e^{-2\pi i q/\tilde{n}},
 \end{equation}
 for any $\mathbf{s} \in \mathbb{R}^2$ we may evaluate $\mathrm{Var}(g(\mathbf{s}))$ as 
 \begin{equation}
     \mathrm{Var}(g(\mathbf{s})) = \sum_{p=1}^{\tilde{n}}\sum_{q=1}^{\tilde{n}}\tilde{f}_{p,q} e^{2\pi i s_1p/\tilde{n}}e^{2\pi i s_2q/\tilde{n}} \qquad \text{ where } \mathbf{s} = (s_1,s_2).
     \label{eq:Fourier_reconstruction}
 \end{equation}
 Thus, we may use \eqref{eq:Fourier_reconstruction} to evaluate $\mathrm{Var}(g(\mathbf{s}_{i,j}))$ {\em without solving the associated linear system}. Instead of evaluating \eqref{eq:Fourier_reconstruction} directly, we use a standard construction whereby the $\{\tilde{f}_{p,q}\}_{p,q=1}^{\tilde{n}}$ is zero-padded to obtain $\{f_{i,j}\}_{i,j=1}^{n}$ as:
\begin{equation}
    f_{i,j} = \begin{cases} 
      \tilde{f}_{p,q} & \text{if } 1 \leq p \leq \frac{\tilde{n}}{2}, 1 \leq q \leq \frac{\tilde{n}}{2} \text{ and } i = p, j = q \\
      \tilde{f}_{p,q} & \text{if } 1 \leq p \leq \frac{\tilde{n}}{2}, \frac{\tilde{n}}{2} < q \leq \tilde{n} \text{ and } i = p, j = q + n - \tilde{n} \\
      \tilde{f}_{p,q} & \text{if } \frac{\tilde{n}}{2} < p \leq \tilde{n}, 1 \leq q \leq \frac{\tilde{n}}{2} \text{ and } i = p + n - \tilde{n}, j = q \\
      \tilde{f}_{p,q} & \text{if } \frac{\tilde{n}}{2} < p \leq \tilde{n}, \frac{\tilde{n}}{2} < q \leq \tilde{n} \text{ and } i = p + n - \tilde{n}, j = q + n - \tilde{n} \\
      0 & \text{otherwise}
    \end{cases}
    \label{eq:zero-padding}
\end{equation}
whence the inverse DFT is applied to $\{f_{i,j}\}_{i,j=1}^{n}$, yielding $\{\mathrm{Var}(g(\mathbf{s}_{i,j}))\}_{i,j=1}^{n}$. We use the ubiquitous Cooley-Tukey Fast Fourier Transform (FFT) \cite{cooley1965algorithm} to compute the DFT and inverse DFT, at a cost of $\mathcal{O}(\tilde{n}\log \tilde{n})$ and $\mathcal{O}(n\log n)$ operations respectively. Thus, Fourier interpolation reduces the number of linear system solves required to compute $\{\mathrm{Var}(g(\mathbf{s}_{i,j}))\}_{i,j=1}^{n}$ from $n^2$ to $\tilde{n}^2$ with negligible computational overhead.

\let\AND\relax 
\begin{algorithm}[t]
\caption{Fourier Interpolation}
\label{alg:fourier_interp}
\begin{algorithmic}[1]
    \STATE{{\bf Inputs:} Parent function $\varphi$ and fine evaluation grid $\{\mathbf{s}_{i,j}\}_{i,j=1}^{n}$ }
    \STATE Choose basis function centers $\{\mathbf{u}_{k,l}\}_{k,l=1}^{r}$ and coarse evaluation grid $\{\mathbf{s}_{i_{p},j_q}\}_{p,q=1}^{\tilde{n}}$ such that \edits{$\tilde{n} \geq 2r + 1$} and $n = M(\tilde{n}-1) + 1$. 
    \STATE{Compute $\displaystyle \{\mathrm{Var}(g(\mathbf{s}_{i_{p},j_q}))\}_{p,q=1}^{\tilde{n}}$ exactly by solving \eqref{eq:exact_variance_solve} for all $\mathbf{s} = \mathbf{s}_{i_{p},j_q}$.}
    \STATE{Compute $\displaystyle \{\tilde{f}_{p,q}\} = \mathcal{F}\left(\{\mathrm{Var}(g(\mathbf{s}_{i_{p},j_q}))\}\right)$}
    \STATE{Compute $\displaystyle \{f_{i,j}\}_{i,j=1}^{n^*}$ by zero-padding, as per \eqref{eq:zero-padding}, where $n^* = M\tilde{n}$.}
    \STATE{Compute $\displaystyle \{\widehat{\text{Var}}_{i,j}\}_{i,j=1}^{n^*} = \mathcal{F}^{-1}\left(\{f_{i,j}\}_{i,j=1}^{n^*}\right)$.}
    \STATE{Compute $\displaystyle \{\widehat{\text{Var}}_{i,j}\}_{i,j=1}^{n}$ by removing the last $M-1$ rows and columns of $\displaystyle \{\widehat{\text{Var}}_{i,j}\}_{i,j=1}^{n^*}$.}
    \STATE{{\bf Return:} $\displaystyle \{\widehat{\text{Var}}_{i,j}\}_{i,j=1}^{n} \approx \{\mathrm{Var}(g(\mathbf{s}_{i,j}))\}_{i,j=1}^{n}$.}
\end{algorithmic}
\end{algorithm}

\begin{figure}[h]
    \centering
    \includegraphics[width = \textwidth]{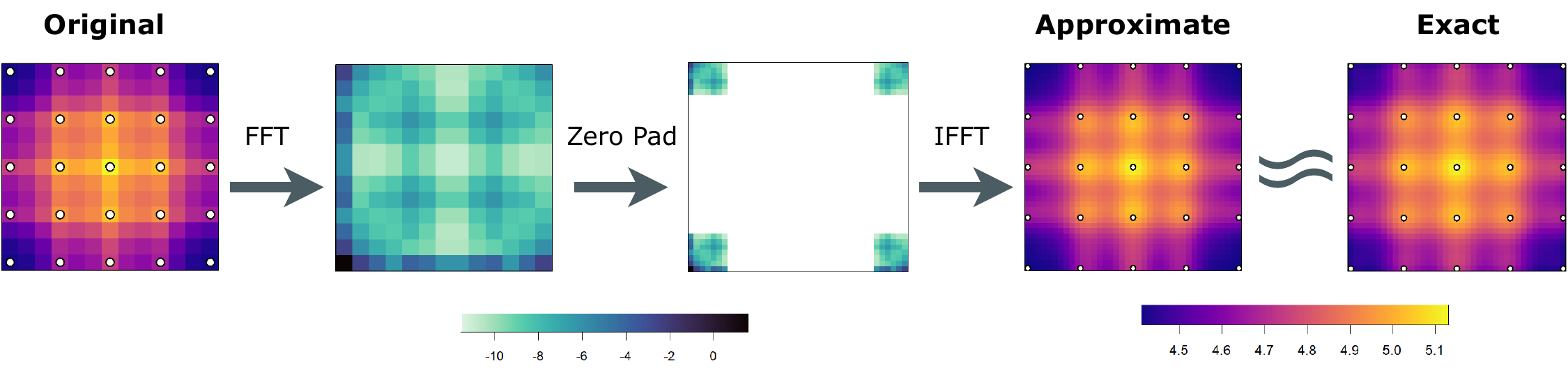}
    \caption{\small Diagrammatic description of the proposed method. In this case $r = 5$ and $\tilde{n} = 13$, with basis function centers denoted by white points. The calculation is then upsampled to a grid with $n = 31$. A visual comparison of the true (exact) variance on the finer grid is provided. The log of the values is taken in the FFT domain to provide a better visual demonstration.}
    \label{fig:method-flowchart}
\end{figure}

In the realistic setting where $\mathrm{Var}(g(\mathbf{s}))$ is only approximately periodic, Fourier interpolation yields an approximation to $\{\mathrm{Var}(g(\mathbf{s}_{i,j}))\}_{i,j=1}^{n}$. Nonetheless, as verified by our experiments in Section~\ref{sec:FFT_Error}, this approximation is in most cases quite good. \edits{While increasing $\tilde{n}$ beyond the theoretical minimum of $\tilde{n}=2r+1$ improves accuracy, it also raises the computational expense. In practice, we find sampling rates of $\tilde{n} = 4r-3$ or $\tilde{n} = 4r$ yield a good balance between these two competing factors.} To avoid errors resulting from content bleeding from the left edge to the right (and the top edge to the bottom) we choose $\tilde{n}$ such that $n = M(\tilde{n}-1) + 1$, for positive integer scale factor $M$, and ensure that the corners of the coarse grid coincide with those of the fine grid. In scenarios such as timing the method, where content bleeding is not a concern, simple modification allows for any choice of $\tilde{n} < n$ to be made. The resulting algorithm is represented pictorially in Figure~\ref{fig:method-flowchart} and in pseudocode as Algorithm~\ref{alg:fourier_interp}, where we use $\mathcal{F}$ to denote the operation of applying the FFT.

\subsection{Normalization using Kronecker Structure}
\label{sec:Kronecker}
An alternate approach to that proposed in Section~\ref{sec:FFT_method}, where the number of linear system solves is reduced, one may also consider speeding up each linear system solve. Elliptic PDEs such as the screened Poisson equation are well-studied, and many fast schemes exist for solving their discretizations \eqref{eq:exact_variance_solve}. One approach utilizes the Kronecker product to represent operations along each dimension separately \cite{chen2016kronecker}. Here, the computation for the variance is exact, but requires constant $\kappa^2$, meaning all diagonal elements of $B$ are the same. 

With reference to \eqref{eq:exact_variance_solve}, we note that $\mathbf{v} \in \mathbb{R}^{R}$ can naturally be reshaped into a matrix $V \in \mathbb{R}^{r\times r}$. Using Kronecker product identities and the ``vec'' vectorization operation
\begin{equation}
    B\mathbf{v} = \operatorname{vec}(AV+ VA^{\top})=\operatorname{vec}(AV I_r+I_r V A^{\top}) = (A \otimes I_r+I_r \otimes A) \operatorname{vec}(V) = (A \otimes I_r+I_r \otimes A) \mathbf{v}
\end{equation}
where $A \in \mathbb{R}^{r\times r}$ is the tridiagonal matrix 
\begin{equation}
A = \left[ \begin{array}{ccccc}
2 + \kappa^2 / 2 & -1 & 0 & \ldots & 0 \\
-1 & 2 + \kappa^2 / 2 & -1 & \ldots & 0 \\
\vdots & & \ddots &  & \vdots \\
0 & \ldots & -1 & 2 + \kappa^2 / 2 & -1 \\
0 & \ldots & 0 & -1 & 2 + \kappa^2 / 2 \\
\end{array} \right]
\end{equation} 
Pre-multiplication by $A$ represents column operations, while post-multiplying by $A$ operates on rows, formulating a discretization across two dimensions. In other words,
\begin{align}
    B = L + \kappa^2I &= A\otimes I_r + I_r\otimes A \label{eq:kronecker}
\end{align}
Equation \eqref{eq:kronecker} may be further simplified using the eigen-decomposition $A = UDU^{\top}$:
\begin{equation}
    B = (U D U^{\top} \otimes I_r+I_r \otimes U D U^{\top})=\mathcal{U D} \mathcal{U}^{\top}
\end{equation}
with $\mathcal{U}=U \otimes U$ and $\mathcal{D}=(D \otimes I_r+I_r \otimes D)$. It also follows that $\mathcal{U}$ is an orthonormal matrix and $\mathcal{D}$ is diagonal, so we have the eigen decomposition for $B$. We use this decomposition to get a compact expression of the process variance:
\begin{equation}
\text{Var}(g(\mathbf{s})) = \boldsymbol{\phi}_{\mathbf{s}}^{\top} Q^{-1} \boldsymbol{\phi}_{\mathbf{s}} = \boldsymbol{\phi}_{\mathbf{s}}^{\top} B^{-1} B^{-\top} \boldsymbol{\phi}_{\mathbf{s}} = \lVert B^{-\top} \boldsymbol{\phi}_{\mathbf{s}} \rVert ^2 = \lVert \mathcal{U} \mathcal{D}^{-1} \mathcal{U}^{\top} \boldsymbol{\phi}_{\mathbf{s}} \rVert^2 = \lVert \mathcal{D}^{-1} \mathcal{U}^{\top} \boldsymbol{\phi}_{\mathbf{s}} \rVert^2
\end{equation}
where the final inequality holds as $\mathcal{U}$ is orthogonal. As $\mathcal{U}^{\top} = U^{\top}\otimes U^{\top}$, we observe that
\begin{equation}
    \mathcal{U}^{\top} \boldsymbol{\phi}_{\mathbf{s}} = \mathrm{vec}\left(U^{\top}\Phi_s + \Phi_s U^{\top}\right)
\end{equation}
where $\Phi_s \in \mathbb{R}^{r\times r}$ is the reshaping of $\boldsymbol{\phi}_{\mathbf{s}} \in \mathbb{R}^{R}$ into matrix form. Consequently, $\mathcal{U}^{\top} \boldsymbol{\phi}_{\mathbf{s}}$ may be computed using only two $r\times r$ matrix-matrix multiplies at a cost of $2(2r^{3} - r^2)$. As $\mathcal{D}$ is diagonal, multiplication by $\mathcal{D}^{-1}$ requires $r^2$ operations. Thus finding the variance at all $N$ spatial locations costs
\begin{equation}
    N\left(4r^3 - 2r^2 + r^2\right) = \mathcal{O}(Nr^3) = \mathcal{O}(NR^{3/2}),
\end{equation}
This method is particularly powerful in settings where the FFT method is limited, such as when the number of basis functions approaches or exceeds the number of points. Timing for this method is provided in Section ~\ref{sec:Timing}. More sophisticated numerical methods for solving elliptic PDEs, such as the multigrid method \cite{multigrid} may yield even greater savings; we leave this for future work. 

\subsection{Multi-Resolution Implementation}
\label{sec:multires}

To guarantee stationarity in a multi-resolution model, each layer of basis functions must be normalized. In order to integrate both the FFT and Kronecker speedups into the {\tt LatticeKrig} R package, two new options were created. The user can choose to use either the Kronecker algorithm alone, or a combined FFT and Kronecker approach. When normalizing a multi-resolution model, the combined approach uses the FFT method for coarser levels, where the number of spatial locations significantly exceeds the number of points, while the Kronecker algorithm is used for the remaining levels. When appropriate, the Kronecker algorithm is also utilized in the inner workings of the FFT method, where it is used to perform the small, exact calculation. This use of both speedups is necessary to avoid hindering the full capabilities of LatticeKrig, as the framework was designed for settings where one can choose $R > N$. Although the bulk of computational overhead lies in the normalization of levels with finer resolutions, the combined approach still demonstrates a significant speedup in Section~\ref{sec:Application}. 

\section{Results}
\label{sec:Results}
We present timing results to demonstrate the computational gains from both algorithms, and assess the consequent approximation errors of the FFT normalization. Our experiments were conducted on a laptop with an 11th Gen Intel(R) Core(TM) i9-11900H processor at 2.50GHz, 16 GB RAM, and 8 cores. For simplicity, all experiments are performed with regular, square grids for both basis functions $\{\mathbf{u}_{k,l}\}_{k,l=1}^r$, and spatial locations $\{\mathbf{s}_{i,j}\}_{i,j=1}^{n}$, with a spatial domain $[-1,1] \times [-1,1]$. Here the coarse grid $\{\mathbf{s}_{i_{p},j_q}\}_{p,q=1}^{\tilde{n}}$ will be chosen with $\tilde{n} = 4r$\edits{, providing a balance between computational efficiency and approximation accuracy.}

\subsection{Timing}
\label{sec:Timing}
Our first aim is to measure the wall clock time required to perform the basis function computation, specifically using the {\tt LKrig.basis} function. We evaluate four distinct scenarios:
\begin{enumerate}
    \item computation without any normalization,
    \item computation with exact normalization using sparse linear algebra (default),
    \item computation leveraging Kronecker products for exact normalization, and
    \item computation using the approximate FFT normalization method.
\end{enumerate}

These experiments contextualize the speedup from both algorithms by recording the time to compute the basis functions; a step that often requires the most computational overhead in a realistic spatial analysis. For each scenario, $r$ is varied between $25$ and $100$, and the side length of the fine grid $n$ ranges from $500$ to $2,000$. Furthermore, we have the option to adjust the $\kappa^2$ parameter and the number of buffer points. For these experiments, we set $\kappa^2 = 0.05$ and employ 10 buffer points in our analysis to mitigate edge effects associated with a finite grid. These choices result in the largest dataset having size $N = 4,000,000$ and a maximum of $R = 10,000$ total basis functions, which are consistent with testing spatial models where traditional, dense linear algebra is not feasible. These results also provide an opportunity to understand how timing will scale for much larger problems. 

\begin{figure}[h]
\centering
\includegraphics[width = 0.85\textwidth]{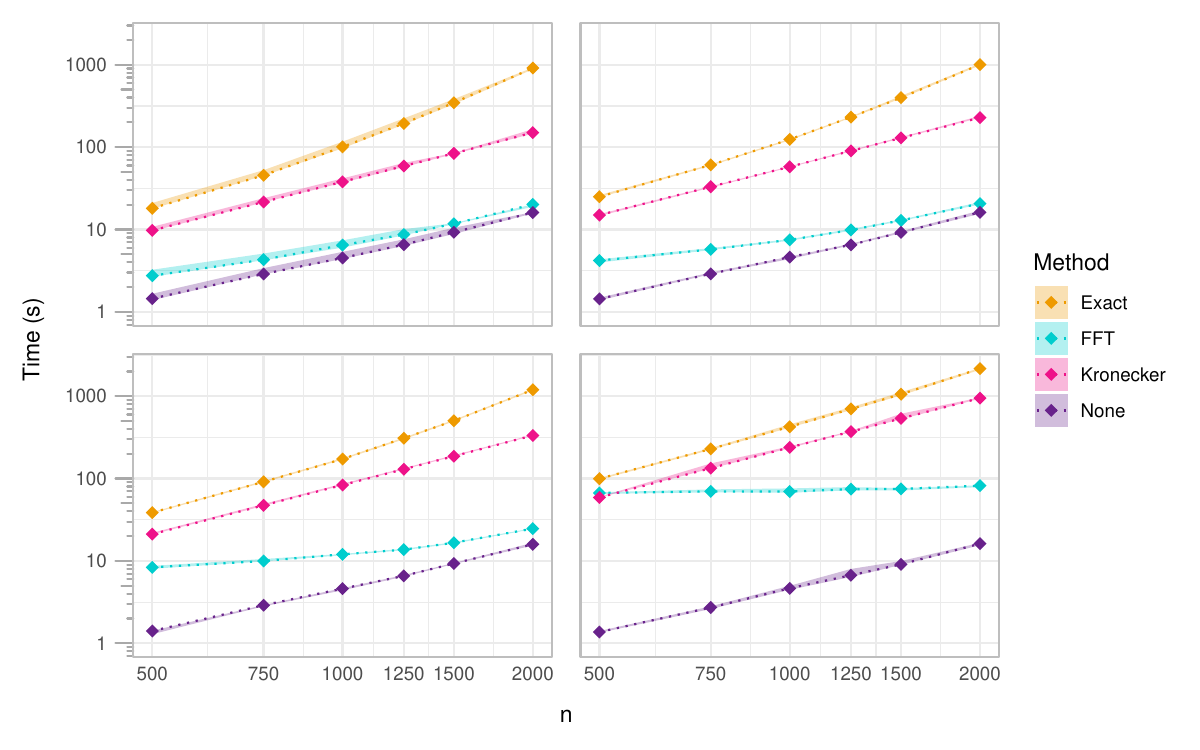}
\caption{\small Timing results on a logarithmic scale for each method where $r$ is varied: 25 (\textbf{top left}), 35 (\textbf{top right}), 50 (\textbf{bottom left}), 100 (\textbf{bottom right}), as is $n$: (500, 750, 1000, 1250, 1500, 2000). Dots represent the median time of 5 iterations of the simulation, and shading represents maximum and minimum values.} 
\label{fig:All_Timing}
\end{figure}

As depicted in Figure \ref{fig:All_Timing}, the FFT normalization algorithm is consistently the fastest, often requiring only marginally more time than the un-normalized calculation. The FFT method achieves maximum efficiency when $r=100$, yielding a speedup factor of $56\times$ faster than the default method. As expected, the timing slope of the FFT method flattens as the number of basis functions increases. This is because the algorithm is primarily limited by the initial, exact calculation on the coarse grid, which is determined by the number of basis functions. Conversely, the Kronecker method emerges as an alternative when accuracy cannot be compromised, offering a speedup of at most $8\times$ over the default method. To roughly illustrate scale between the three normalization methods, we take the case with $r=50$ and $n=2000$. The default method requires approximately 20 minutes, the Kronecker algorithm takes about 5 minutes, and the FFT algorithm takes a mere 25 seconds.

\subsection{Error in the FFT Method}
\label{sec:FFT_Error}
\begin{figure}[h]
\centering
\includegraphics[width = 0.7\textwidth]{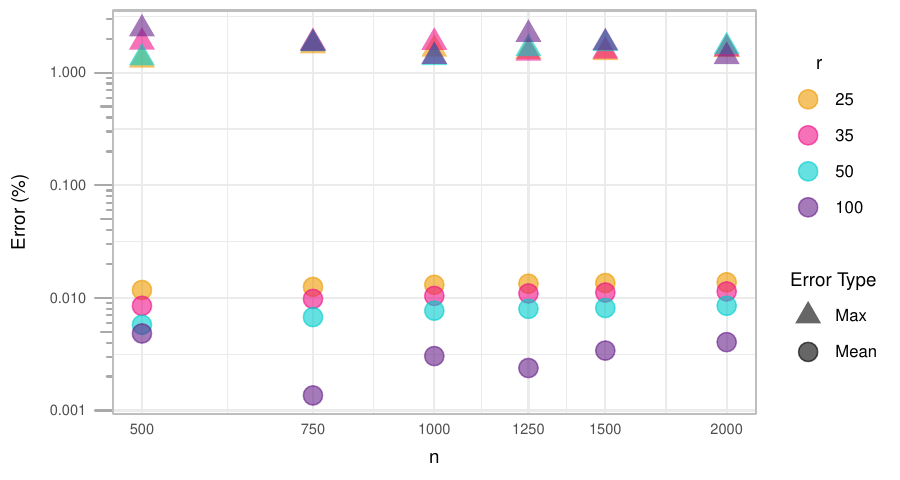}
\caption{\small Mean and maximum error on a logarithmic scale for the FFT normalization method. Once again, both $r$: (25, 35, 50, 100), and $n$: (500, 750, 1000, 1250, 1500, 2000) are varied. } 
\label{fig:All_Error}
\end{figure}

It is important to determine the approximation error from the FFT-based method. Given the lack of physical significance attached to the variances, we use relative error metrics for comparison. Figure \ref{fig:All_Error} shows both the mean percentage error and maximum absolute percentage errors, computed by subtracting the FFT approximation $\{\widehat{\mathrm{Var}}(g(\mathbf{s}_{i,j})\}_{i,j}$ from the results of the exact calculation at each $\mathbf{s}_{i,j}$. Across all tested configurations of basis function counts and grid side lengths, we find that the error introduced by the FFT method remains consistently low. Specifically, the mean percent error hovers around 0.01\%, while the maximum values range between 1\% and 2\%. We demonstrate that the impact of this error is essentially negligible in Section 5, when prediction accuracy and presence of artifacts are evaluated.

Additionally, we find these results are robust to different values of $\kappa^2$ and variations in the number of buffer points. Error is minimized at the default value of $\kappa^2 = 0.05$, with the maximum error increasing to roughly $4.5\%$ at a large value of $\kappa^2 = 1$. A choice of 10 buffer points---slightly higher than the default value of 5---minimizes the error, although all tested values between (and including) $5$ and $40$ do not surpass a maximum error of $3\%$. When the number of buffer points decreases to 0, error rises as high as 35\%, as the assumption of periodicity is strongly violated in this scenario. Edge effects due to finite grids are a well known issue, rendering the high error from a lack of buffer points irrelevant, as this scenario would not be seen in practice.

\section{Application}
\label{sec:Application}

\begin{figure}[H] 
\centering
\includegraphics[width = 0.87\textwidth]{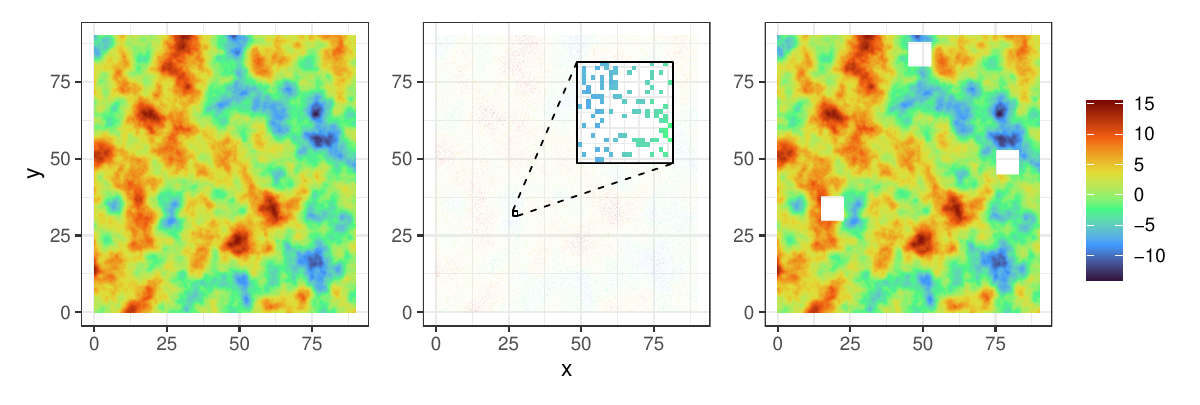}
\caption{\small Full data set of $N = 1,329,409$ locations (\textbf{left}) along with training visualizations of both the MAR (\textbf{center}) and Blocks (\textbf{right}) sampling schemes.} 
\label{fig:simulated_train_true}
\end{figure}

We perform a comparison of prediction results and timing using the different normalization methods on a large, simulated dataset. Circulant embedding \cite{circulantembed} is used to simulate a Gaussian field on a $1153 \times 1153$ grid with a spatial domain of $[0,90] \times [0,90]$, a Matern covariance parameterized by a range of $\theta =6$ and a smoothness $\nu = 1$, and added white noise $\varepsilon \sim \mathcal{N}(0, 0.2)$. Observation values $Z(\mathbf{s})$ are scaled using an affine transformation that results in a range of roughly $-14.05$ to $15.45$. We employ two sampling schemes, shown in Figure \ref{fig:simulated_train_true}. The first is Missing at Random (MAR), in which 80\% of the values are randomly removed, leaving the remaining 20\% as training data. The second is Missing Blocks (Blocks), in which we remove three square blocks of size $100 \times 100$---in total removing $30,000$ observations from the image---and test the models' long range predictive abilities. 
\looseness=-1
 
\noindent We record the run time of the full spatial analysis, and generate prediction surfaces to assess the accuracy of out-of-sample predictions relative to the test data for each of the four normalization choices: 

\begin{enumerate} 
    \item None: No normalization at all. 
    \item Both: The combined algorithm, using the FFT at coarser layers and the Kronecker for the rest. 
    \item Kronecker: The exact Kronecker algorithm. 
    \item Exact: The default, exact computation. 
\end{enumerate}

We fit a LatticeKrig model with 4 levels such that $r$ is 25, 49, 97, and 193, respectively ($R = 65,844$ total basis functions). When the ``both'' speedup is used in this experiment, FFT normalization occurs on the first 3 levels, and the last level uses the Kronecker speedup. We choose a value of $\kappa^2 = 0.015$ and a buffer of $10$ basis functions, following the findings in Section~\ref{sec:FFT_Error}. The times for both fitting the model and generating a prediction surface on the whole grid are combined to form total run time. 

\begin{table}[H] 
\centering
\begin{tabular}{lllllllll}
\hline
\multirow{2}{*}{Method} & \multicolumn{4}{c}{Blocks}          & \multicolumn{4}{c}{MAR}       \\ \cline{2-9} 
                        & MAE $\downarrow$  & RMSPE $\downarrow$ & Time (min) & & MAE $\downarrow$ & RMSPE $\downarrow$ & Time (min) & \\ \hline
None                    & 1.1672 & 1.6432 & 28.68 & & 0.1983 & 0.2486 & 16.57 &  \\
Both                    & 1.0527 & 1.4735 & 78.82 & & 0.2051 & 0.2570 & 45.08 &  \\
Kronecker               & 1.0508 & 1.4712 & 111.49 & & 0.2051 & 0.2570 & 57.90 &  \\
Exact                   & 1.0508 & 1.4712 & 173.75 & & 0.2051 & 0.2570 & 95.51 &  \\ \hline
\end{tabular}
\caption{\small Mean absolute error (MAE), root-mean-squared prediction error (RMSPE), and total run time are recorded for each choice of normalization method for both the MAR and Blocks sampling schemes. Arrows ($\uparrow, \downarrow$) indicate the desirable direction for a metric.}
\label{prediction-table}
\end{table}

\begin{figure}[H] 
\centering
\includegraphics[width = 0.9\textwidth]{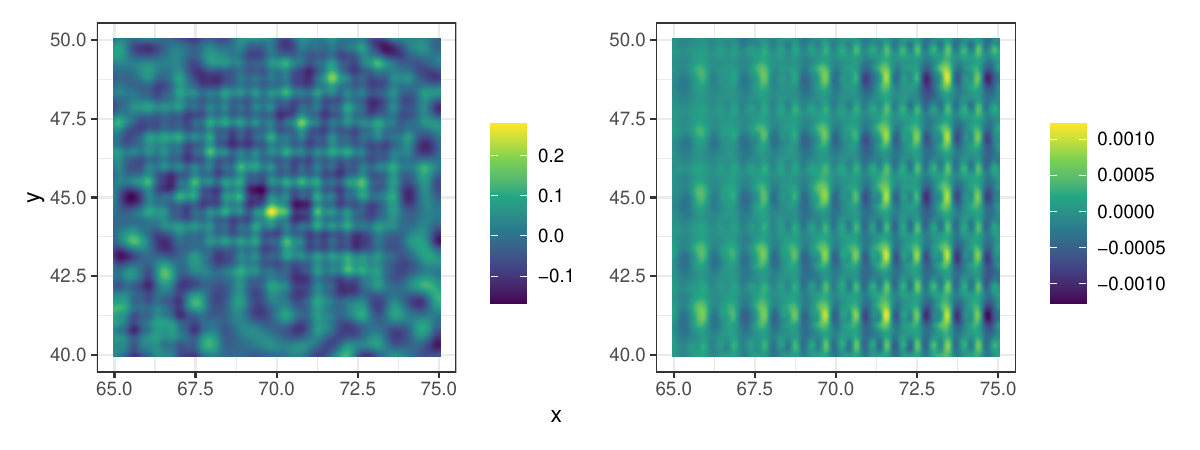}
\caption{\small Artifacts obtained by taking the difference of the unnormalized and exactly normalized prediction surfaces in the MAR case (\textbf{left}). The remnants of those artifacts after approximate normalization (\textbf{right}).} 
\label{fig:simulated_artifacts}
\end{figure}

As seen in Table \ref{prediction-table}, the ``Kronecker'' and ``Both'' options offer significant speedups in the context of the full spatial analysis. The combined option is roughly 25\% faster than the strictly Kronecker-based counterpart, and roughly $2\times$ faster than the classic normalization method. We expect time differences between the three normalization methods to continue to widen as data-set size increases due to their differing computational complexities. The Blocks case takes more time as the number of points one needs to fit the model to initially is much greater than the MAR case. As for accuracy, all choices of normalization provide similarly excellent fit. 
The impact from the artifacts in the un-normalized case is first noticeable in the differing values of MAE and RMSPE, while the results from the approximate, combined approach are near identical to those of the exact methods. In the Blocks case, the un-normalized method is less accurate, while in the MAR case it performs slightly better; an unexpected result. Despite this observation, we do not deem these differences in accuracy significant, as the true detriment of artifacts comes from their inaccurate representation of physical data. This is clearly demonstrated in Figure \ref{fig:simulated_artifacts}, which zooms in to a small subset of the data (lat: $[40,50]$, lon: $[65,75])$ to display the artifacts more clearly.

The periodic pattern from not normalizing is striking, and is reminiscent of the basis functions and the shape of their variance. The average amplitude of the unnormalized artifacts is on the order of $10^{-2}$. The combined approach removes roughly 99\% of the artifacts, reducing their average amplitude to an order of $10^{-4}$. If scaled to match the range of the no-normalization image (left), the image on the right side of Figure \ref{fig:simulated_artifacts} displays as a solid color due to the greatly diminished amplitude. Note that both the default and Kronecker normalization methods would remove the artifacts entirely. 

\section{Conclusion}
\label{sec:Conclusion}

A useful model for analyzing large spatial data is the representation of a Gaussian Process as a sum of basis functions with a multivariate normal distribution prescribed for the coefficients. An important adjustment to this model is a normalization to avoid artifacts in the predicted surface that arise from the arrangement of the locally supported basis functions. Despite the value from eliminating these artificial features, this normalization step constitutes a computational bottleneck. In this work we have introduced two fast algorithms for this step, reducing the cost of estimation and prediction for large spatial data sets. These methods also constrain the spatial model to have a constant marginal variance, another attractive feature.

The Kronecker method takes advantage of the SAR structure of the model's precision matrix, providing a way to speed up each individual linear system solve required for normalization. This is an exact calculation and enjoys a speedup of roughly $5\times$ over the default method for computing normalization weights. However, it is limited to the case of a constant $\kappa^2$ and a two dimensional, rectangular geometry. In addition, even if these conditions are met, there exist faster ways of solving discretizations of elliptic PDEs, which could be utilized for this purpose. A future research direction is to leverage numerical and graphical algorithms for PDEs to improve the computation for the normalization, or to derive approximate but analytical expressions for the variance function. 

An alternative to the Kronecker approach, the FFT method requires fewer linear systems solves, and instead upsamples the variance to much finer grids. Although this method is approximate, it is very accurate, and can be faster than the default by almost two orders of magnitude. Additionally, it can be used for any non-stationary covariance model for the coefficients, and could be expanded to work for higher dimensions.  The FFT method is currently confined to models or layers where the number of basis functions is significantly less than that of the number of points, and requires the coarse, exact calculation to be done on a nested subgrid of the data. There exists a broad range of other image upsampling techniques that could be implemented, using our general methodology as a guide. In particular, the self-similarity of the variance function could be exploited to avoid many exact evaluations.

To streamline the examples in this work, we have assumed that the data locations are registered to a grid. This restriction can be overcome by computing the normalization on a fine grid, on the order of the data location spacing, and then using a fast local interpolation for the off-grid locations. Because the FFT method can interpolate to very fine grids efficiently, handling irregularly spaced locations will not impact the timing.

All numerical experiments for this paper were done specifically within the LatticeKrig framework. This was done for the convenience of working within a tested R package and also to facilitate distribution of the algorithms. However, provided the radial basis functions are organized on a regular grid and the coefficients follow a sparse precision matrix, these algorithms could be useful for other basis function models. Future work could study other basis functions that are more amenable to normalization, such as tensor products, and also geometries where rectangular grids are not available, such as spheres. In general, we believe that basis function models are an important tool for large spatial data analysis, and fine tuning their properties to match standard spatial models is an important area for further research.

\section{Acknowledgments} 
The authors acknowledge the work of Nathan Lenssen in implementing the Kronecker normalization method into the {\tt LatticeKrig} package. We would also like to thank Brennan Sprinkle for helpful discussion. Finally, we are grateful to the reviewers and editors for their comments and suggestions that improved the quality of this manuscript. 

\section{Conflict of Interest Statement}
The authors declare no potential conflict of interests.

\section{Data Availability Statement}
The data that support the findings of this study are openly available in Normalization-Paper at \\ \href{https://github.com/antonyxsik/Normalization-Paper}{\tt github.com/antonyxsik/Normalization-Paper}. 

\printbibliography

\appendix

\section{Further details on solving the screened Poisson equation}
\label{appendix_a}
Recall that the equation we wish to solve is 
\begin{equation}
    -\Delta v_{\mathbf{s}} +\kappa^2 v_{\mathbf{s}} = \psi_{\mathbf{s}}(\mathbf{z}). 
    \label{eq:screened_poisson_appendix}
\end{equation}
Assuming $v_{\mathbf{s}}$ (respectively $\psi_{\mathbf{s}}$) is absolutely integrable, it may be expressed in terms of its Fourier transforms, $\tilde{v}_{\mathbf{s}}$ (resp. $\tilde{\psi}_{\mathbf{s}}$):
\begin{equation}
    v_{\mathbf{s}}(\mathbf{z}) = \iint_{\mathbb{R}^2} e^{2\pi i \mathbf{z}\cdot\boldsymbol{\omega}} \tilde{v}_{\mathbf{s}}(\boldsymbol{\omega})d\boldsymbol{\omega} \qquad \text{ and } \qquad \psi_{\mathbf{s}}(\mathbf{z}) = \iint_{\mathbb{R}^2} e^{2\pi i \mathbf{z}\cdot\boldsymbol{\omega}} \tilde{\psi}_{\mathbf{s}}(\boldsymbol{\omega})d\boldsymbol{\omega}.
\end{equation}
We substitute these expressions into \eqref{eq:screened_poisson_appendix}, transforming the differential equation to an algebraic one in the frequency domain:
\begin{align}
            & \|\boldsymbol{\omega}\|^2 \tilde{v}_{\mathbf{s}} + \kappa^2\tilde{v}_{\mathbf{s}} = \tilde{\psi}_{\mathbf{s}}. \\
\Rightarrow \ & \tilde{v}_{\mathbf{s}} = \frac{1}{\kappa^2 + \|\boldsymbol{\omega}\|^2} \tilde{\psi}_{\mathbf{s}} \\
\Rightarrow \ & v_{\mathbf{s}}(\mathbf{z}) = \iint_{\mathbb{R}^2}\frac{e^{-2\pi i\mathbf{z}\cdot\boldsymbol{\omega}}}{\kappa^2 + \|\boldsymbol{\omega}\|^2}\tilde{\psi}_{\mathbf{s}}(\boldsymbol{\omega})d\boldsymbol{\omega} \label{eq:Fourier_solution} 
\end{align}
by the ``shift'' property of Fourier transforms:
\begin{equation}
    \tilde{\psi_{\mathbf{s}}}(\boldsymbol{\omega}) := \iint_{\mathbb{R}^2}\psi_{\mathbf{s}}(\mathbf{z})e^{2\pi i\mathbf{z}\cdot\boldsymbol{\omega}} d\mathbf{z} = \iint_{\mathbb{R}^2}\psi(\|\mathbf{z} - \mathbf{s}\|)e^{2\pi i\mathbf{z}\cdot\boldsymbol{\omega}} d\mathbf{z} = e^{2\pi i\mathbf{s}\cdot\boldsymbol{\omega}}\tilde{\psi}(\boldsymbol{\omega})
\end{equation}
where $\tilde{\psi}(\cdot)$ is the Fourier transform of $\psi(\cdot)$.
Returning to \eqref{eq:Fourier_solution}, we may rewrite the solution as
\begin{equation}
    v_{\mathbf{s}}(\mathbf{z}) = \iint_{\mathbb{R}^2}\frac{e^{-2\pi i(\mathbf{s} - \mathbf{z})\cdot\boldsymbol{\omega}}}{\kappa^2 + \|\boldsymbol{\omega}\|^2}\tilde{\psi}(\boldsymbol{\omega})d\boldsymbol{\omega}
\end{equation}

\section{Software Integration}
In order to set up a model in {\tt LatticeKrig}, one typically creates an {\tt LKrig} object, in which they define critical parameters such as $\kappa^2$, number of buffer points, overlap, and normalization choice.

One can choose between all three normalization options by simply modifying:
{\tt LKrigSetup(..., normalize = TRUE, normalizeMethod = "option",...)}. 
The options for the normalization method are ``exact'', ``exactKronecker'', or ``both''. The ``exact'' option selects the default, slow normalization method, and the ``exactKronecker`` option selects the Kronecker method. These will be used for each layer within a multi-resolution model. The ``both'' option uses the combined approach, as described in Section~\ref{sec:multires}. Setting {\tt normalize = FALSE} makes all options irrelevant, and does not normalize the basis functions. The implementation in {\tt LatticeKrig} provides defaults for this strategy, along with examples and help files to illustrate software details. The development version can be found on \href{https://github.com/antonyxsik/Normalization-Paper}{\tt github.com/antonyxsik/Normalization-Paper}.

\end{document}